# Drive and damping govern resonant electron energy redistribution in a kinetic model informed by helium discharges

Zhao-Ju Bo[1,2], Lei Chang[1,2*], Shi-Jie Zhang[1], Ying-Xin Zhao[1], Zi-Chen Kan[1], Ji-Kai Sun[1], Ilya Zadiriev[3], Elena Kralkina[3], Shin-Jae You[4], Jing-Jing Ma[1], Husi Géza[5]

[1]School of Electrical Engineering, Chongqing University, Chongqing 400044, China
[2]Institute of Nuclear Energy and Technology Innovation, Chongqing University, Chongqing 400044, China
[3]Physical Electronics Department, Faculty of Physics, Lomonosov Moscow State University, GSP-1, Leninskie Gory, Moscow, 119991, Russian Federation
[4]Applied Physics lab for PLasma Engineering (APPLE), Department of Physics, Chungnam National University, Daejeon 34134, Republic of Korea
[5]Institute of Vehicle Engineering and Mechatronics, Faculty of Engineering, University of Debrecen, Ótemető Strt. 2-4, Debrecen 4028, Hungary

[*]Email: leichang@cqu.edu.cn, leichang.plasma@gmail.com

## Abstract

Energetic electrons exchange energy with electrostatic waves in low-temperature plasmas, but wave amplitude alone does not determine the direction or cumulative effect of that exchange. We analyse nonlinear electron redistribution using a single-mode kinetic model parameterized by published helium-discharge measurements. Frequency and pressure comparisons are combined with a controlled damping scan and explicit particle-current and energy diagnostics. Across the frequency cases, the inferred resonant drive varies by a factor of 2.7 while background damping varies by only about 9%. Reducing pressure from 33.3 to 10.66 Pa at 37 MHz increases the net drive by a factor of 2.24. In the damping scan, the peak wave energy and residual parallel-energy deficit both decrease by approximately 93%, consistent with quadratic field-amplitude scaling at the reported endpoints. An integrated energy identity connects retained wave energy, background loss and distribution recovery, and distinguishes cessation of wave growth from reversal of energy transfer. Quantitative timescale and directional-distribution tests define the connection between this reduced model and discharge kinetics. The analysis provides a common interpretation of resonant redistribution, coherent current and energy exchange, together with observables for testing that interpretation in kinetic simulations and phase-resolved measurements.

# Introduction

Electron energy distributions control excitation, ionization and reactive-species production in low-temperature plasmas [1–4]. In capacitively coupled plasmas (CCPs), moving sheaths accelerate electrons into the bulk, where collective fields can redistribute their energy [5–8]. The resulting electron response depends on both the directional distribution and the phase relation between current and field. Resolving how those quantities change during a nonlinear wave event is therefore important for relating plasma operating conditions to energy deposition.

Particle-in-cell simulations with Monte Carlo collisions resolve electron motion together with the radiofrequency (RF) driven sheaths, self-consistent fields, collisions and particle boundaries. Such calculations have already established collective electron dynamics and energetic-beam formation in capacitive discharges [3,6–8]. These phenomena motivate a local wave–particle analysis, while their identification as a particular instability requires a directional distribution and the relevant kinetic dispersion relation. Published energy distributions supply useful input trends.

The bump-on-tail (BOT) instability provides a controlled framework for separating resonant energy supply, wave damping and recovery of a driving distribution. Its linear growth, trapping and plateau formation are established plasma physics [9–17]. The question addressed here is how changes in these competing rates appear jointly in the wave amplitude, residual electron-energy change and current–field power transfer. Comparing those diagnostics on consistent energy scales is particularly useful when operating conditions change several input parameters simultaneously.

We examine this question using a single-mode Berk–Breizman (BB) model with input families informed by helium-discharge measurements at 10 W of dissipated power [18]. The measurements span 13.56 MHz–43 MHz at 33.3 Pa and two pressures at 37 MHz. A separate damping scan holds the remaining model inputs fixed. The combined analysis identifies the rates organizing the nonlinear response, establishes an energy-consistent comparison of the scan endpoints, and derives a history-dependent balance linking electron redistribution to wave storage and loss. Explicit boundary conditions and a quantitative applicability map make the physical content of this local model testable against more complete discharge descriptions.

# Results

## Discharge-informed families of resonant inputs

The calculations represent a periodic, locally uniform interaction between a selected electrostatic wave and energetic electrons. Bulk electrons supply the cold-fluid response and a prescribed damping rate $\gamma_d$; energetic electrons supply the resonant drive $\gamma_L$ and relax toward a fixed target at rate $\beta$. RF frequency and gas pressure label the input families. The applied RF field, electrodes and reaction network are outside this local initial-value problem; their influence enters through the prescribed distributions and rates. The model equations and reconstruction conventions are given in Methods. The numerical benchmark is shown in Supplementary Fig. S1.

Figure 1 shows the directional surrogate distributions used to select the resonance and local positive gradient. Their density-preserving mapping from published electron energy probability functions (EEPFs) imposes a directional interpretation, which is stated explicitly in Methods. The main dynamical controls are the damping ratio $d = \gamma_d/\gamma_L$ and recovery ratio $b = \beta/\gamma_L$. The quantity $\kappa = \beta/(\gamma_L - \gamma_d)$, retained in the original parameter tables, is related by $b = \kappa(1 - d)$. Table 1 summarizes the input families; Supplementary Tables S1–S3 give the representative case, frequency scan and pressure scan.

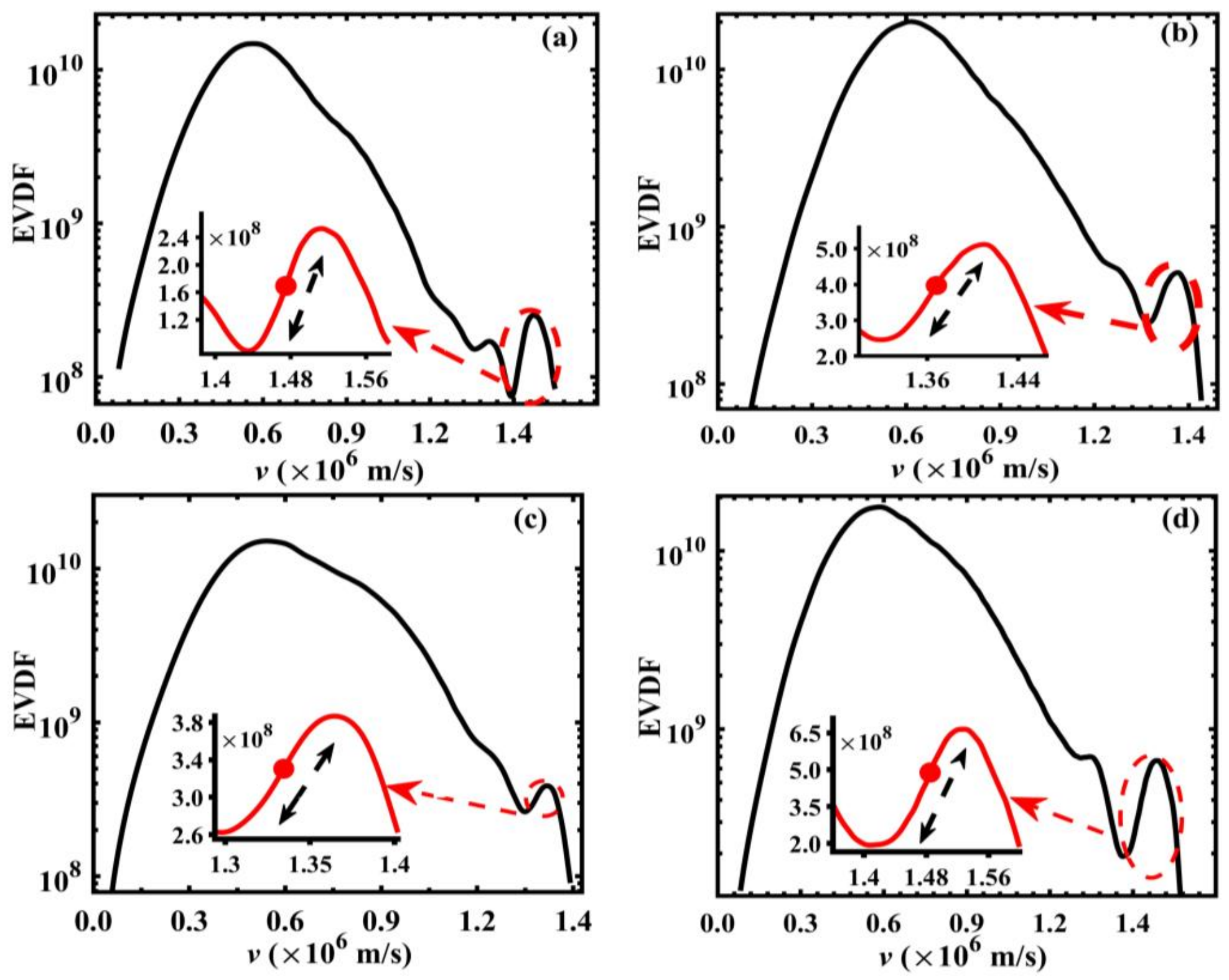

Figure 1. Equivalent one-dimensional EVDFs constructed from the measured EEPFs at (a) 13.56 MHz, (b) 27 MHz, (c) 37 MHz, and (d) 43 MHz. The insets show the positive-slope intervals used as model inputs; dashed lines delimit the gradient fits, and red markers indicate the selected resonant velocities.

Table 1. Input families for the reduced model. Plasma frequency and linear growth rate are in $10^9$ s$^{-1}$. The five cases use $D = S = 0$. The recovery parameter $\kappa = \beta/(\gamma_L - \gamma_d)$; original parameter tables are retained in Supplementary Tables S1–S3.

| RF (MHz) | Pressure (Pa) | $\omega_{pe}$($10^9$ s$^{-1}$) | $\gamma_L$($10^9$ s$^{-1}$) | $\gamma_d/\gamma_L$ | $\kappa$ |
|---|---|---|---|---|---|
| 13.56 | 33.3 | 7.03 | 3.740 | 0.071 | $1.03 \times 10^{-4}$ |
| 27 | 33.3 | 7.88 | 3.145 | 0.090 | $4.62 \times 10^{-5}$ |
| 37 | 33.3 | 8.00 | 2.070 | 0.134 | $2.31 \times 10^{-4}$ |
| 43 | 33.3 | 8.28 | 5.600 | 0.052 | $1.19 \times 10^{-4}$ |
| 37 | 10.66 | 8.43 | 4.110 | 0.023 | $2.49 \times 10^{-5}$ |

With a constant equilibrium slope, $D = S = 0$ and fixed normalized initial and numerical conditions, the evolution is specified by $d$ and $b$. Input families sharing these quantities have the same normalized dynamics. Frequency and pressure select points in this parameter space and set the dimensional conversion scales. This similarity makes the rate comparison transferable to other directed populations described by the same local equations.

## Frequency dependence is organized mainly by the resonant drive

The inferred drive exceeds the background damping at all four frequencies. From Supplementary Table S2, the net linear rates $\gamma_L - \gamma_d$ are $3.474\times 10^9$ s$^{-1}$, $2.862\times 10^9$ s$^{-1}$, $1.793\times 10^9$ s$^{-1}$, and $5.310 \times 10^9$ s$^{-1}$ at 13.56 MHz, 27 MHz, 37 MHz, and 43 MHz, respectively. The dimensional damping rate varies by only about 9% across the set, while the linear drive varies by a factor of 2.7. The distribution-based parameterization thus identifies variation of the resonant drive as the main source of the frequency-dependent response. Its combination with a relatively narrow range of dimensional damping produces an appreciable change in $\gamma_d/\gamma_L$, which governs the subsequent normalized evolution.

Figure 2 shows growth followed by oscillatory decay. The largest peak and longest persistence in normalized time occur for the 43 MHz input family, and the weakest response occurs at 37 MHz. Their respective damping ratios are 0.052 and 0.134. Thus the frequency ordering is explained by the rates supplied to the resonance model, rather than by identifying the RF excitation itself with the electrostatic mode frequency. Frequency-dependent sheath heating at fixed power supplies the discharge context for these input changes [18,19]. Supplementary Figs. S2 and S3 show the associated resonant distributions and phase-space evolution.

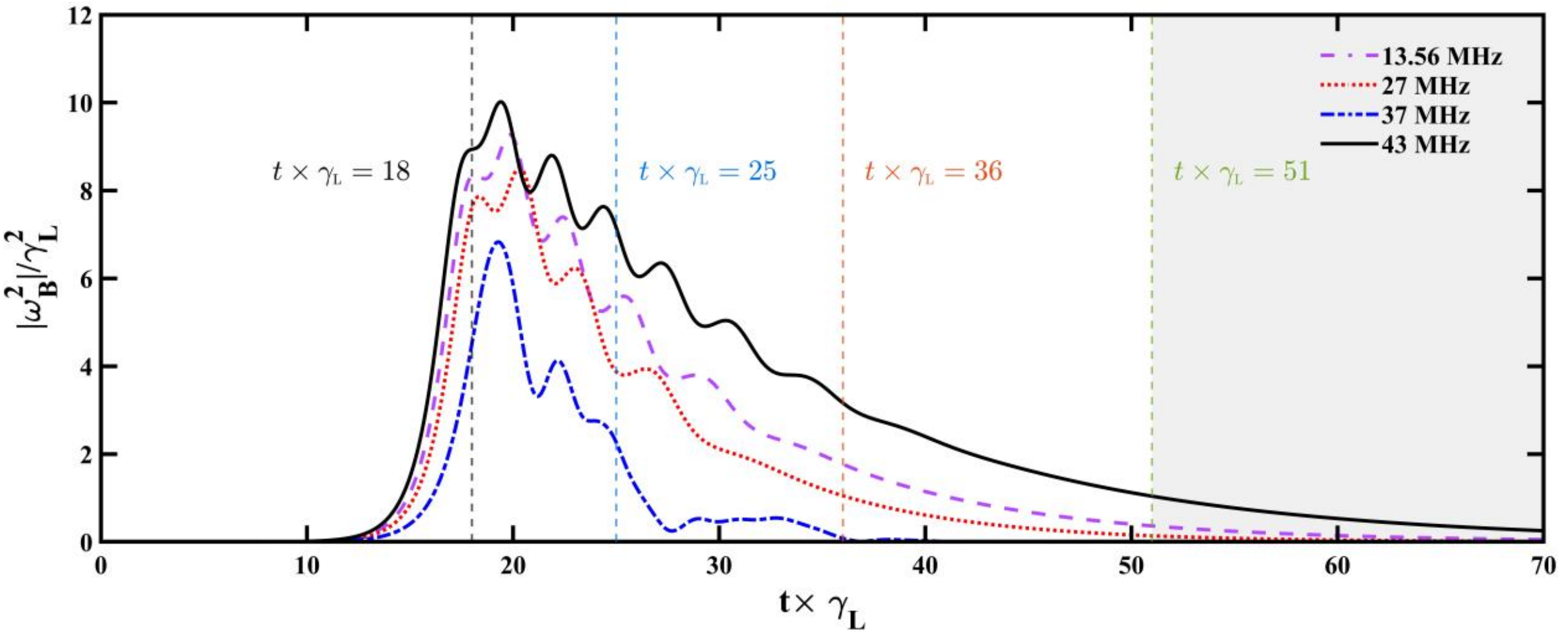

Figure 2. Normalized squared bounce frequency $|C| = \omega_B^2/\gamma_L^2$ versus $\tau = \mathrm{t} \times \gamma_L$ for the four RF frequencies at 33.3 Pa. The dashed lines mark the 37 MHz snapshots in Supplementary Fig. S3. Each curve uses its own linear-drive rate for normalization.

The resonant distributions and phase-space snapshots in Supplementary Figs. S2 and S3 connect the envelope to trapping, filamentation and detrapping. The variation in bounce period across trapped orbits shears the distribution into fine filaments. Accumulation below resonance and depletion above it reduce the initial positive slope and weaken the resonant drive. As the wave potential weakens, the trapping region contracts and releases electrons into passing orbits. The spatially averaged profile remains altered after the coherent island fades.

The mean parallel-energy perturbation $\delta T$ is the second-moment diagnostic defined in Methods. Figure 3 shows energy extraction during wave growth, followed by partial return and damped oscillations. The oscillations reflect repeated energy exchange between the wave and electrons trapped in its potential wells. Background damping removes part of the wave energy during this exchange, reducing the energy available for return to the electrons. At $\tau = 100$, Supplementary Fig. S4 shows residuals of approximately $-1.80 \times 10^{-3}$ eV, $-1.70 \times 10^{-3}$ eV, $-0.95 \times 10^{-3}$ eV and $-2.03 \times 10^{-3}$ eV for 13.56 MHz, 27 MHz, 37 MHz and 43 MHz, respectively. The magnitudes of these residual deficits follow the ordering of the net drives.

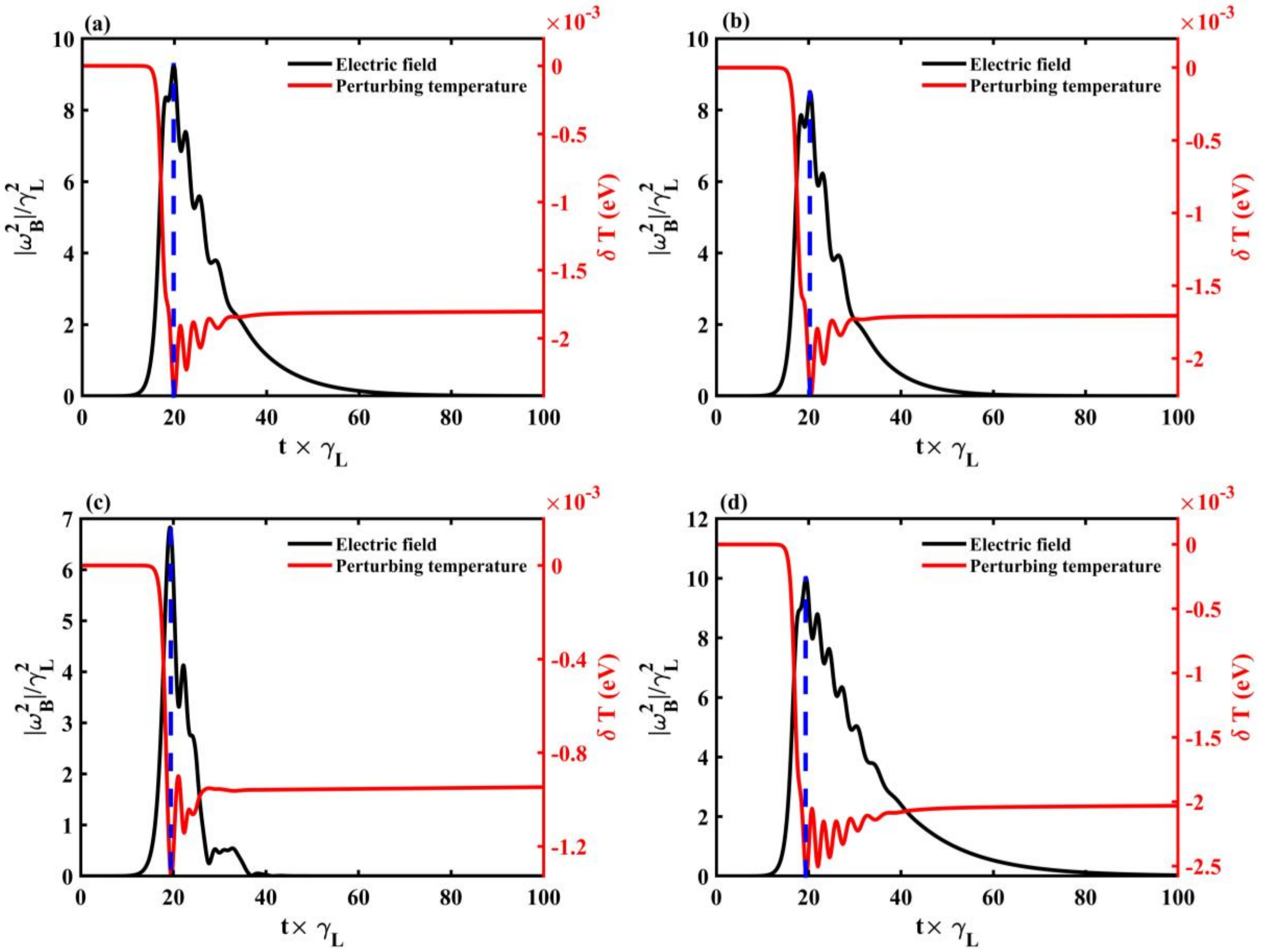

Figure 3. Evolution of the normalized squared bounce frequency and mean parallel energy perturbation $\delta T$ of the modeled energetic-electron population at (a) 13.56 MHz, (b) 27 MHz, (c) 37 MHz, and (d) 43 MHz. The horizontal coordinate is $\tau = t\gamma_L$.

Time is normalized separately in each case: $t = \tau/\gamma_L$. The four $\tau = 100$ samples therefore correspond to 26.7 ns, 31.8 ns, 48.3 ns and 17.9 ns. Likewise, a plotted width $\Delta u$ in $u = (kv - \omega)/(2\pi\gamma_L)$ corresponds to $\Delta v = (2\pi\gamma_L/k)\Delta u$. These conversions distinguish the dimensionless rate comparison from dimensional predictions for the participating electrons. The separatrix half-width is $2\omega_B/k$, so the normalized trapping width follows $\sqrt{|\mathrm{C}|}$.

## Pressure changes reinforce the balance between drive and loss

Reducing the pressure from 33.3 to 10.66 Pa changes the plasma frequency by only about 5%, from $8.00 \times 10^9\ \mathrm{s}^{-1}$ to $8.43 \times 10^9\ \mathrm{s}^{-1}$, while nearly doubling the inferred drive from $2.07 \times 10^9\ \mathrm{s}^{-1}$ to $4.11 \times 10^9\ \mathrm{s}^{-1}$. The damping simultaneously decreases from $2.77 \times 10^8\ \mathrm{s}^{-1}$ to $0.94 \times 10^8\ \mathrm{s}^{-1}$. Consequently, $\gamma_d/\gamma_L$ falls from 0.134 to 0.023, and $\gamma_L - \gamma_d$ increases from approximately $1.79 \times 10^9\ \mathrm{s}^{-1}$ to $4.016 \times 10^9\ \mathrm{s}^{-1}$, a factor of about 2.24. The pressure response is thus governed by a substantial rebalancing of drive and loss despite the modest change in plasma frequency.

The input distributions used for this comparison are shown in Supplementary Fig. S5. The low-pressure case has both a stronger inferred drive and weaker background loss. It develops a larger and more persistent normalized mode amplitude (Supplementary Fig. S6) and a broader affected interval in the resonant distribution (Supplementary Fig. S7). Its mean parallel-energy perturbation is more negative and evolves to later normalized times (Fig. 4) than the corresponding 33.3 Pa result (Fig. 3(c)).

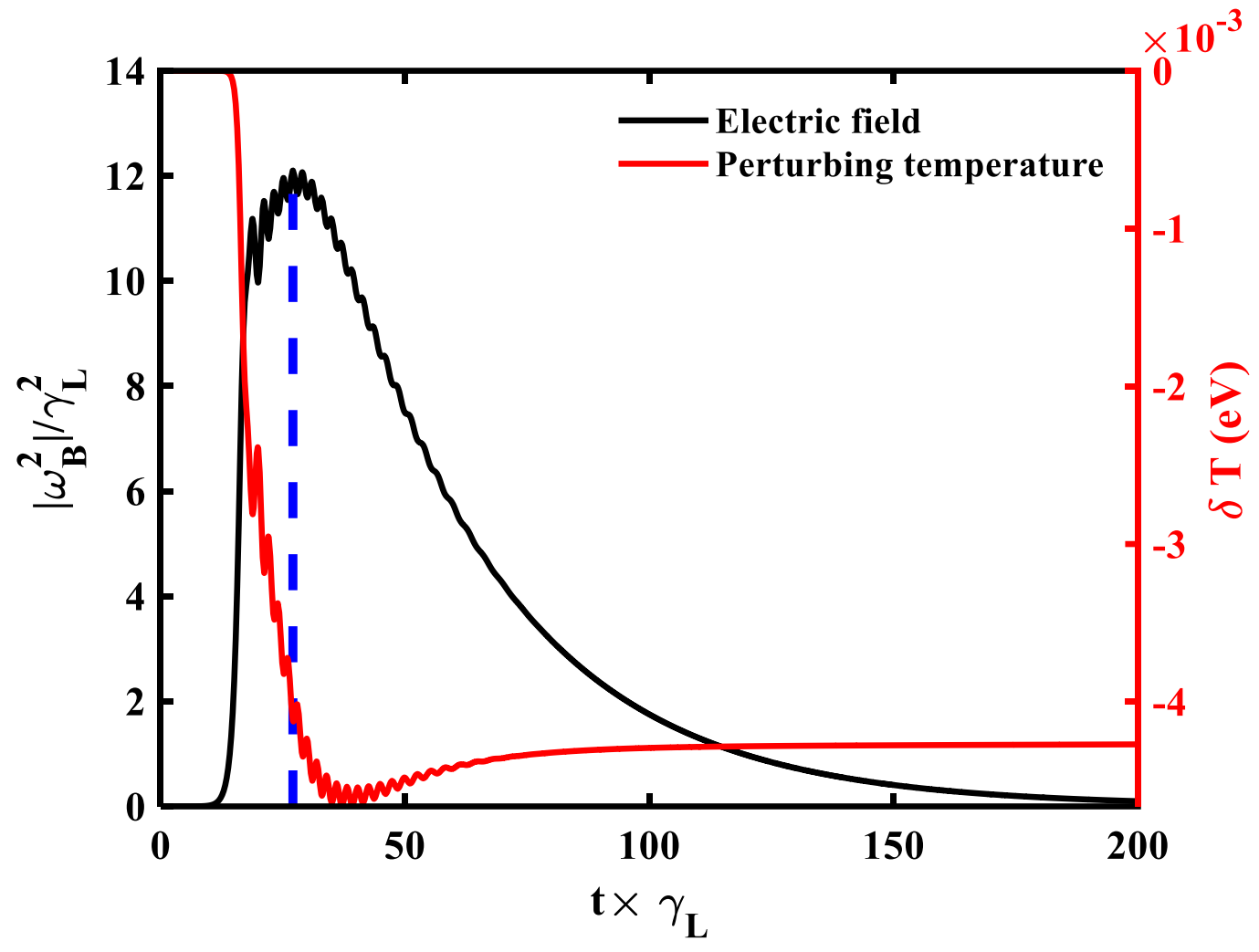

Figure 4. Normalized squared bounce frequency and change in mean parallel electron energy at 10.66 Pa and 37 MHz; the 33.3 Pa counterpart is shown in Fig. 3(c).

The pressure comparison also separates redistribution from recovery. At $\tau = 124$, $\beta t$ is approximately 0.0030 at 10.66 Pa and 0.025 at 33.3 Pa. Direct Krook relaxation changes a fixed perturbation by at most about 2.5% over this interval. The larger response at low pressure is therefore organized primarily by drive, wave damping and nonlinear trapping, while restoration of the initial target acts on a longer timescale. The two $\tau = 124$ profiles correspond to 30.2 ns and 59.9 ns, respectively.

## Current and field resolve the direction of energy exchange

The first distribution harmonic gives a coherent electron flux and conduction current whose wavelength averages vanish. Their phase relation to the field determines the nonzero mean power transfer. The relation $\delta J_e = -e\,\delta\Gamma_e$ expresses the coherent response in electrical and particle units. The reconstructed current at a fixed wave phase is provided in Supplementary Fig. S8, and the harmonic conventions are specified in Methods.

The phase-averaged power changes sign with the current–field correlation. Positive power denotes energy transferred from the mode to electrons; negative power denotes electron drive. In the BOT envelope, the retained resonant current replaces $v$ by $v_r$ inside the first moment. Methods distinguishes that approximation from the full velocity-weighted current and derives the finite-width correction to the combined energy identity.

Figure 5 resolves this exchange together with the mode-energy variation and the non-negative damping loss. During growth, $-p_e$ exceeds the background loss $p_d$. At a wave-energy maximum, the two are equal: the energetic electrons continue to supply energy even though the envelope has stopped growing. The field can begin to decay before the resonant power reverses sign. A later positive $p_e$ identifies energy return through the retained resonant channel. This temporal distinction explains why the electron-energy minimum need not coincide exactly with the field maximum.

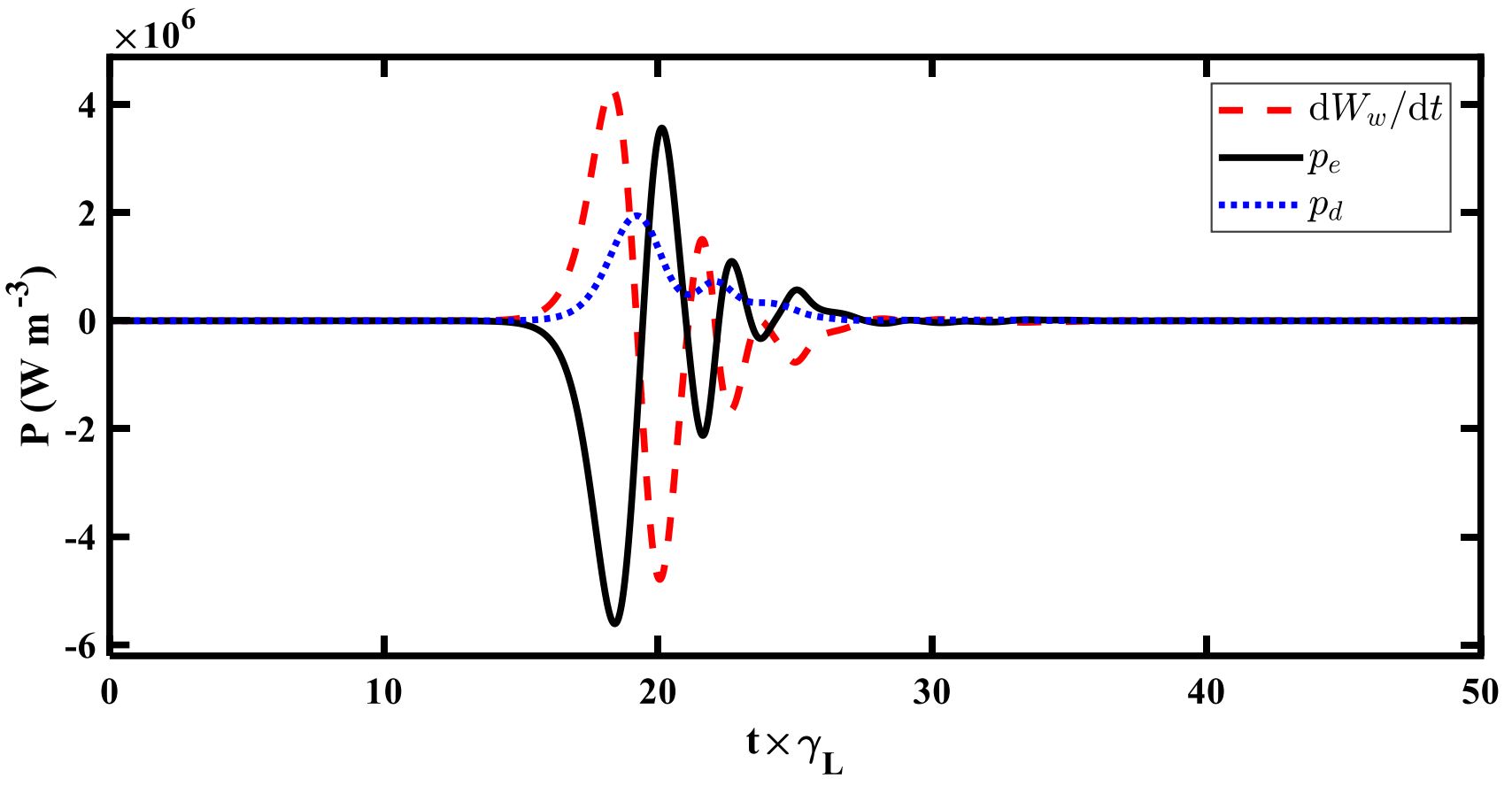

Figure 5. Phase-averaged mode-energy variation $dW_w/dt$, resonant-electron power $p_e$, and background-damping loss $p_d$ at 33.3 Pa and 37 MHz. Negative $p_e$ denotes electron drive of the wave.

The integrated balance gives a direct interpretation of the residual change. For a consistently reconstructed, fixed-density energetic population, define $Y = |C|^2$ and $H$ as its energy perturbation divided by the mode-energy scale. Methods derives an exponentially weighted identity for $H + Y$, including the correction between the full and retained resonant currents. Within the matched resonant approximation and weak-recovery limit, the electron-energy deficit equals retained mode energy plus cumulative background loss, after subtracting the seed energy. The identity connects the waveform to the energy moment and distinguishes a physical finite-width correction from a numerical balance error.

## The damping scan gives consistent changes on the two energy scales

With the remaining inputs fixed, increasing $\gamma_d/\gamma_L$ from 0.002 to 0.2 reduces the peak normalized amplitude $|C|$ from about 22.0 to 5.67. The residual $|\delta T|$ decreases from $1.07 \times 10^{-2}$ eV to $7.15 \times 10^{-4}$ eV (Fig. 6). The field-amplitude reduction is about 74%, and the corresponding peak trapping-width ratio is $\sqrt{5.67/22.0} \approx 0.51$.

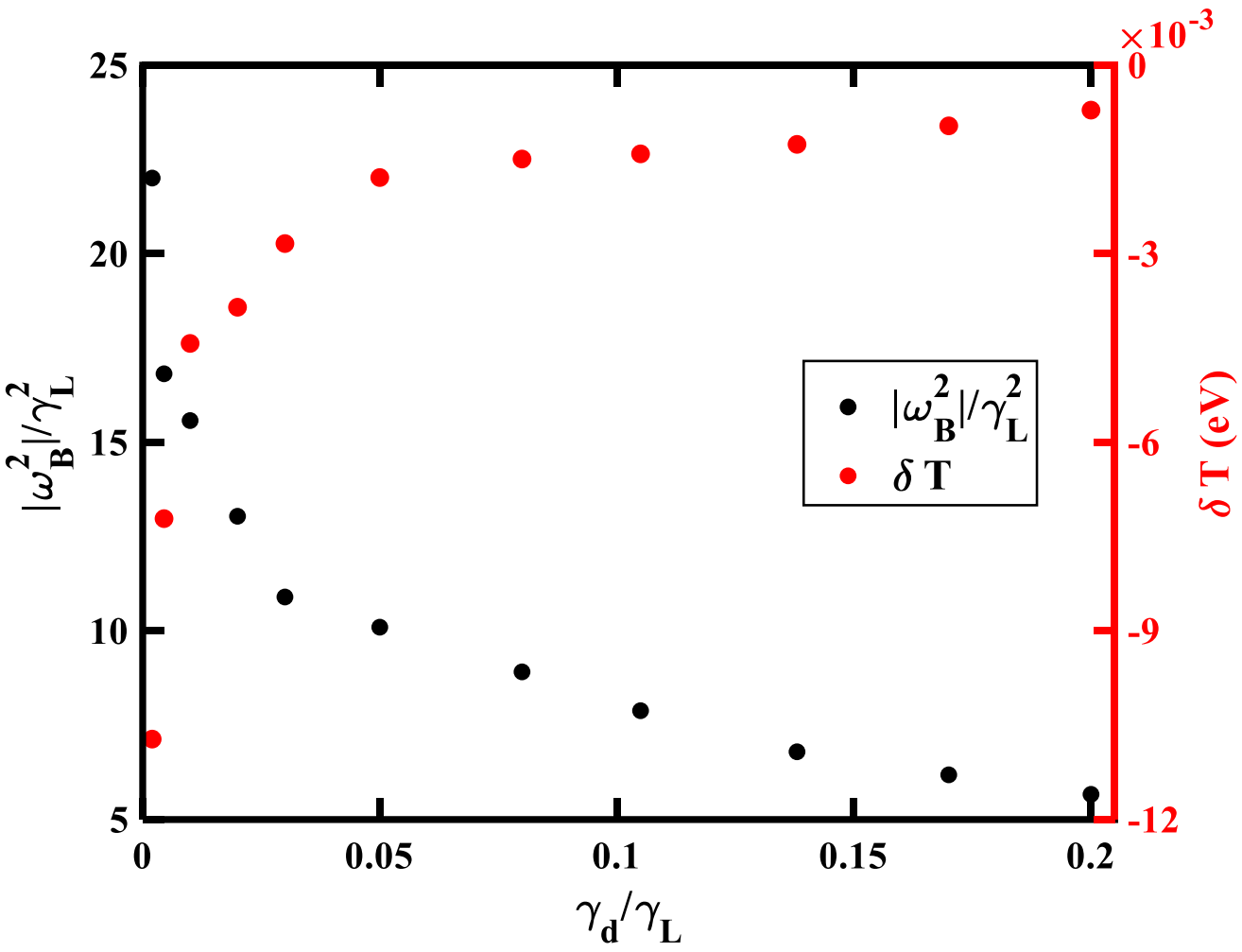

Figure 6. Peak normalized squared bounce frequency and late-time mean parallel electron-energy perturbation versus $\gamma_d/\gamma_L$, with all other model inputs held fixed.

Wave energy is proportional to $|C|^2$, so the ratio of the endpoint peak wave energies is $(5.67/22.0)^2 \approx 0.0664$. The residual electron-energy ratio is $(7.15 \times 10^{-4})/(1.07 \times 10^{-2}) \approx 0.0668$. Both energy measures therefore decrease by approximately 93%, with a relative difference of about 0.6% between the ratios calculated from the rounded endpoints. This is a consistency of energy scaling at the reported endpoints. The time histories and integrated energy identity provide the separate information needed to resolve interaction duration, energy storage and return.

The distribution and current diagnostics support the dynamical interpretation (Supplementary Figs. S9 and S10). Stronger damping reduces the coherent response and the interval of sustained trapping. The weaker wave potential narrows the velocity interval accessible to trapping, limiting the range over which trapped electrons can flatten the driving positive slope.

## Discussion

The central result is a common rate and energy description of the nonlinear response. Across the frequency families, the large change in inferred drive dominates the much smaller variation in dimensional damping. Lower pressure strengthens the drive while reducing wave loss. The independent damping scan then isolates the role of prescribed wave loss, while the current–field phase identifies the direction of resonant exchange. The study thereby connects operating-condition-dependent inputs to three complementary observables: resonant redistribution, coherent current and cumulative electron-energy change.

The energy analysis sharpens the physical meaning of saturation. Stopping wave growth requires a balance between continuing electron drive and background loss; returning energy to electrons requires a subsequent reversal of the current–field correlation. The residual moment records the integrated balance over that sequence. Its dependence on the waveform follows from the energy identity with the stated current and boundary terms, while the scan endpoints are already consistent with quadratic amplitude scaling. This separates a supported numerical comparison from the more general history dependence required by energy conservation.

Table 2 makes the separation of timescales explicit. The small accumulated recovery, $\beta t = 0.003\text{–}0.025$, explains persistence of the redistributed profile over the sampled interval. Phase mixing transfers coherent information to fine velocity-space structure and attenuates low-order moments; it is not itself irreversible fine-grained dissipation [20–22]. The explicit damping and Krook terms determine the energy sinks and sources. Late-time profiles here describe the finite observation window.

Table 2. Ordering parameters calculated directly from the input families. $\beta t$ and $t/T_{RF}$ are evaluated at $\tau = 124$; $T_{RF}$ is the period of the applied excitation. These values characterize recovery, mode ordering and the frozen-background interval.

| RF (MHz) | Pressure (Pa) | $\gamma_L/\omega_{pe}$ | $\beta t$ | $t/T_{RF}$ |
|---|---|---|---|---|
| 13.56 | 33.3 | 0.532 | 0.0119 | 0.45 |
| 27 | 33.3 | 0.399 | 0.0052 | 1.06 |
| 37 | 33.3 | 0.259 | 0.0248 | 2.22 |
| 43 | 33.3 | 0.676 | 0.0140 | 0.95 |
| 37 | 10.66 | 0.488 | 0.0030 | 1.12 |

The same map identifies the connection to a physical discharge. The input values $\gamma_L/\omega_{pe} = 0.26\text{–}0.68$ are outside a uniformly weak-growth ordering, and the representative $k\lambda_{De} \approx 0.5$ makes thermal dispersion consequential. A warm-fluid estimate gives $\omega/\omega_{pe} \approx \sqrt{1 + 3k^2\lambda_{De}^2} \approx 1.32$ [1,2]. A kinetic dielectric calculation would additionally determine bulk Landau damping and shift the selected resonance. The present comparisons are thus nonlinear responses of the prescribed reduced equations; their dimensional rates are estimates under that closure. Varying an effective damping rate tests loss sensitivity but cannot reproduce all changes in a warm dispersion relation. The separatrix half-width relative to $v_r$ is $2(\gamma_L/\omega_{pe})\sqrt{|C|}$ and must be compared with the interval over which the target slope is treated as constant. A broad resonant interval also calls for evaluating the full-current correction retained in the energy identity. These spatial and velocity scales complement the time-scale assessment.

The energy distribution determines neither the angular distribution nor the directional slope uniquely. For an isotropic, nonnegative distribution, the exact directional marginal has a non-positive derivative at positive velocity. The driving gradient used here therefore represents an assumed directed energetic population. Sheath-generated beams and high-frequency collective responses established in kinetic studies of CCP studies [3,6–8] provide physical motivation for this assumption, while they are distinct from observing the modeled instability in the helium measurements used for input. Direction-resolved distributions and a positive net growth rate from the full dielectric response would establish that connection.

A complete discharge calculation also evolves the applied RF field, sheaths, wall loss and electron emission, ionization and excitation, and energy-dependent elastic and inelastic scattering. In the local model, the background is frozen and wave damping and distribution recovery are separate prescribed terms. Electron–neutral scattering of resonant electrons may change their direction and detrapping rate as well as attenuate the bulk response; it is not determined solely by the bulk-averaged collision rate. The relevance of a local resonance is set by comparing its growth and trapping times with those processes and with the spatial scale of background variation. At $\tau = 124$ the sampled evolution spans 0.45 RF–2.22 RF periods, so connection of the late-time response to a discharge requires the RF-phase-dependent background.

The framework supplies concrete tests for that connection. A phase-resolved directional distribution determines the sign and width of the resonant gradient; a high-frequency field spectrum determines the mode frequency and wavenumber; simultaneous current and field measurements determine the sign of resonant power. A kinetic discharge simulation can evaluate the same moments and separate energetic-electron exchange from bulk conduction, sheath work and inelastic loss. Agreement among these observables, rather than plateau formation alone, would identify the modeled energy-transfer mechanism in an operating plasma.

The present results establish how discharge-informed input families organize nonlinear resonant redistribution and how to compare the resulting diagnostics consistently. The energy-scale comparison, separation of saturation from energy-return onset, and explicit ordering map provide the principal advances of the analysis. They make the reduced calculation useful as an interpretable component of broader kinetic studies and specify the measurements needed to assess its contribution to electron-energy redistribution.

# Methods

## Local kinetic model and boundary conditions

We resolve the electrode-normal coordinate $x$ and velocity $v = v_x$ in a locally uniform interaction region. Ions provide a stationary neutralizing background on the electron timescale. The bulk electrons are described by a cold collisional fluid, while a dilute energetic population is treated kinetically. The electric field contains a single spatial harmonic; the kinetic distribution retains the harmonics generated by its nonlinear interaction with that field.

The starting kinetic–fluid equations follow the BB formulation [13–17]. We write the dimensional equations using the International System of Units (SI) and take $e > 0$ as the magnitude of the electron charge. Here, $F(x, v, t)$ is the energetic-electron distribution, $V$ is the bulk-electron fluid velocity, and $\nu_c$ is the effective bulk momentum-transfer frequency. The background density $n_e$ enters the cold-electron response:

$$\frac{\partial F}{\partial t} + v\frac{\partial F}{\partial x} - \frac{|e|E}{m_e}\frac{\partial F}{\partial v} = \left.\frac{dF}{dt}\right|_{\text{coll}}, \tag{1}$$

$$\frac{\partial V}{\partial t} = -\frac{|e|E}{m_e} - \nu_c V, \tag{2}$$

$$\frac{\partial E}{\partial t} = \frac{e}{\varepsilon_0}\left[n_e V + \int v(F - F_0)dv\right]. \tag{3}$$

Here, $F_0(v)$ is the unperturbed energetic-electron distribution. The electric field is written as $E(x,t) = \text{Re}\left[E_1(t)e^{i\xi}\right]$, and $F - F_0$ is expanded in harmonics of $\xi = kx - \omega t$. The cold-background approximation sets $\omega \approx \omega_{pe}$ and the resonant velocity $v_r = \omega/k$. Near this velocity, $F_0$ is approximated by a constant positive gradient, from which the linear resonant drive $\gamma_L$ is calculated using Eq. (11). A positive tail gradient supplies resonant drive; net instability additionally requires that this drive exceed the modeled losses. Equation (2) gives the amplitude damping rate $\gamma_d = \nu_c/2$ in the weakly damped cold-plasma limit. Equation (2) supplies the linearized cold-bulk response. In the single-mode reduction, Eq. (3) is projected onto the fundamental spatial harmonic; higher distribution harmonics influence the retained field through nonlinear coupling to the first harmonic. The kinetic source–relaxation operator is prescribed as:

$$\left.\frac{dF}{dt}\right|_{\text{coll}} = D\frac{\partial^2}{\partial v^2}(F - F_0) + S\frac{\partial}{\partial v}(F - F_0) - K(F - F_0). \tag{4}$$

The three terms represent velocity-space diffusion, drag, and relaxation toward $F_0$, with characteristic rates $\nu = (k^2 D)^{1/3}$, $\alpha = (kS)^{1/2}$, and $\beta = K$, respectively. We set $D = S = 0$ to isolate the competition between wave damping and Krook recovery of the resonant distribution. The bulk momentum-loss rate controls coherent wave damping, whereas the Krook rate controls relaxation toward the prescribed kinetic target. The Fourier representation imposes spatial periodicity over one wavelength:

$$F(x + L, v, t) = F(x, v, t),\ E(x + L, t) = E(x, t),\ L = 2\pi/k.$$

The coordinate $x$ therefore labels a local periodic interaction region, rather than a finite electrode gap. The model has no wall reflection, emission or imposed electrode potential. The electrostatic mode is self-consistently driven by the kinetic current in the amplitude equation; the external RF field enters only through the selected input family. Its frequency is distinct from the modeled carrier $\omega \approx \omega_{pe}$. Ions provide the stationary neutralizing background.

The constant-gradient reduction advances perturbations around the local target $F_0$ near resonance. In the whole-velocity formulation, the moment identities below require an integrable physical reconstruction for which the total distribution and the relevant velocity-weighted boundary terms vanish. For a finite reconstruction interval, endpoint particle and energy fluxes must instead be retained. This distinction matters because the local affine target used to derive the spectral solver is not itself a global density-normalized distribution.

## Normalization and parameter construction

The bounce frequency $\omega_B = (|e|k|E_1|/m_e)^{1/2}$ characterizes deeply trapped electrons in the sinusoidal wave potential. The complex amplitude $C = |e|kE_1/(m_e\gamma_L^2)$ retains the field phase, while its magnitude is the squared bounce frequency normalized to the linear drive. Time and detuning are normalized by the same drive rate. With the SI form of the distribution normalization, the definitions are

$$|C| = \omega_B^2/\gamma_L^2, \tag{5}$$

$$G = \frac{e^2\omega_{pe}}{2\varepsilon_0 m_e\gamma_L^2 k}F, \tag{6}$$

$$\tau = \gamma_L t, \tag{7}$$

$$\hat{\gamma}_d = \gamma_d/\gamma_L, \tag{8}$$

$$\Omega = \left(kv - \omega_{pe}\right)/\gamma_L, \tag{9}$$

$$\bar{v} = v/\gamma_L, \tag{10}$$

$$\gamma_L \equiv \left(\omega_{pe}^3\pi/2k^2 n_e\right)(\partial F_0/\partial v). \tag{11}$$

The harmonics $G_n(\Omega,\tau)$ of $F - F_0$ are coupled to $C(\tau)$ through the normalized kinetic and amplitude equations [13–17]. We use the BOT numerical framework [20,23–28] to evolve this system. For $D = S = 0$, its prescribed relaxation and damping parameters are $\beta/\gamma_L$ and $\gamma_d/\gamma_L$. The tables additionally report $\kappa = \beta/(\gamma_L - \gamma_d)$, which compares relaxation with the net linear growth rate. These normalizations are related by $\beta/\gamma_L = \kappa(1 - \gamma_d/\gamma_L)$. The normalization must also be retained when comparing different operating conditions. A common normalized time does not imply a common physical time, and a common detuning interval does not imply a common velocity interval. In dimensional form, $t = \tau/\gamma_L$, $v - v_r = (\gamma_L/k)\Omega$, and $|E_1| = (m_e\gamma_L^2|C|)/(ek)$. The plotted amplitude $|C|$ is proportional to the field amplitude; wave energy is proportional to $|C|^2$.

The measured EEPFs are taken from the helium experiment of Abdel-Fattah [18], which used fixed dissipated power of 10 W. The density-preserving conversion uses the electron energy distribution $F_\varepsilon(\varepsilon) = dN/d\varepsilon$. For the common EEPF convention $q(\varepsilon)$, $F_\varepsilon(\varepsilon) = \sqrt{\varepsilon}q(\varepsilon)$, with $\varepsilon$ expressed in eV. The corresponding speed distribution is $P(v) = (m_e v/e)F_\varepsilon(m_e v^2/2e)$. Rescaling the speed coordinate by $1/\sqrt{3}$ then gives $g(u) = \sqrt{3}P(\sqrt{3}u),\ u > 0$. The distribution and its assigned density refer to the same energetic population. The energy measurements determine the speed distribution $P(v)$ but do not specify the electron angular distribution. The model adopts the rescaled $g(u)$ as the distribution of the velocity component along the wave propagation direction for a directed energetic population and evaluates the resonant slope from this constructed distribution. Figure 1 and Supplementary Fig. S5 identify the selected resonant regions.

For comparison, if $f_{iso}(v)$ is an isotropic three-dimensional velocity distribution, its directional marginal and derivative satisfy $g(u) = 2\pi \int_{|u|}^{\infty} v f_{iso}(v) dv$ and $\frac{dg}{du} = -2\pi u f_{iso}(u) \leq 0$ for $u > 0$. Equal second moments therefore do not reconstruct a positive directional slope. A fitted slope and its fitting interval, the chosen resonance and the density normalization specify the kinetic input.

The angular assumption also has an explicit mathematical realization. Let $P(v)$ be the speed distribution and $h(\mu|v)$ a normalized pitch-angle density, with $\mu = v_x/v$. Their directional marginal is $g(u) = \int_{|u|}^{\infty} \frac{P(v)}{v} h(u/v|v) dv$. The present rescaling $g(u) = \sqrt{3}\, P(\sqrt{3}\, u), \quad u > 0$ is obtained, for example, from a fixed cone $\mu = 1/\sqrt{3}$ with uniform azimuth. This population has equal raw second moments in the three Cartesian components while remaining directional. Supplementary Methods derives the result and distinguishes a single directed population from two opposite populations.

The 37 MHz, 33.3 Pa case illustrates the parameter evaluation. The fitted gradient is $\partial F_0/\partial v = 1867.43\ \mathrm{m^{-5} s^2}$, and the plasma frequency is $\omega = \omega_{pe} = \sqrt{(n_e e^2)/(\varepsilon_0 m_e)} = 8.0 \times 10^9\ \mathrm{s^{-1}}$. Combining the resonance condition $\omega = kv$ with the selected velocity $v = 1.331 \times 10^6\ \mathrm{m\ s^{-1}}$ gives $k = \omega/v$ and $k \approx 6.0 \times 10^3\ \mathrm{m^{-1}}$. The corresponding wavelength is approximately 1.0 mm, with $k\lambda_{De} \approx 0.5$, and Eq. (11) yields $\gamma_L = 2.07 \times 10^9\ \mathrm{s^{-1}}$. This sequence connects the selected resonant region to the characteristic spatial and temporal scales of the modeled instability.

At a neutral-gas temperature of 300 K, $N_g = p/(k_B T_g) = 8.04 \times 10^{21}\ \mathrm{m^{-3}}$. Averaging the electron–neutral momentum-transfer rate over the distribution using $\nu_{en} = \frac{N_g}{n_e} \int_0^{\varepsilon_{max}} \sigma_m(\varepsilon) v(\varepsilon) F_\varepsilon(\varepsilon) d\varepsilon$ gives $\nu_{en} \approx 5.54 \times 10^8\ \mathrm{s^{-1}}$. Here $\int_0^{\varepsilon_{max}} F_\varepsilon(\varepsilon) d\varepsilon = n_e$ normalizes the bulk energy distribution over the represented energy range, $\sigma_m$ is the momentum-transfer cross section and $v(\varepsilon) = \sqrt{2e\varepsilon/m_e}$ for $\varepsilon$ in eV. The bulk distribution used in this average is distinct from the assigned energetic population used to construct the resonant slope. The cold-fluid damping prescription then yields $\gamma_d = \nu_{en}/2 = 2.77 \times 10^8\ \mathrm{s^{-1}}$ and $\gamma_d/\gamma_L = 0.134$. The ratio of plasma to neutral density is approximately $2.5 \times 10^{-6}$. Electron–neutral momentum transfer supplies the explicit background-loss mechanism, while the slower Coulomb energy exchange enters through the relaxation estimate below. This separation allows wave dissipation and recovery of the energetic-electron distribution to be assessed independently.

The Krook term relaxes the energetic-electron distribution toward $F_0$ at the rate $\beta = 1/\tau_c$. Its dual role is to damp perturbations and restore the prescribed driving gradient. The Coulomb energy-exchange estimate in Ref. [29] supplies the characteristic equilibration time used for this effective relaxation model:

$$\tau_c = \frac{3\sqrt{2m_e}\pi^{3/2}\epsilon_0^2}{n_0 e^4 \ln \Lambda} (T_{\mathrm{bulk}} + T_{\mathrm{EP}})^{3/2}. \tag{12}$$

Equation (12) estimates the equilibration time of two Maxwellian electron populations [29]. Here $n_0$ is the background-electron density used for the equilibration estimate and $\ln\Lambda$ is the Coulomb logarithm. Its SI prefactor requires $T_{bulk}$ and $T_{EP}$ to be expressed as energies in joules; temperatures quoted in eV are multiplied by $e$ before evaluation. For the non-Maxwellian energetic tail, this estimate supplies the effective energy-exchange scale used to specify the Krook rate.

The effective Krook rate is fixed by the tabulated $\kappa$ values using $\beta/\gamma_L = \kappa(1-\gamma_d/\gamma_L)$. This fully specifies recovery in the normalized calculations. Equation (12) supplies a two-Maxwellian energy-exchange estimate for interpreting that parameter; its use for a non-Maxwellian tail is an effective relaxation model.

## Spectral evolution and numerical verification

The BOT formulation [20,23–28] expands the distribution perturbation in spatial harmonics and Fourier transforms velocity detuning to its conjugate coordinate $s$. The characteristic velocity is $ds/d\tau = -n$ for harmonic $n$. The published BOT implementation uses a finite conjugate-velocity interval with zero incoming perturbation at the positive boundary and outgoing information discarded at the negative boundary. The shift is zero padded, not wrapped periodically. Spatial periodicity and this open spectral boundary are therefore separate conditions. The characteristic update, source integration and nonlinear iteration are given in Supplementary Methods.

In the published seed construction, all distribution-perturbation harmonics initially vanish and the electric-field amplitude provides the perturbation. The representative parameter table specifies $\omega_B(0)/\gamma_L = 10^{-3}$, corresponding to $|C(0)| = 10^{-6}$. A uniform spatial translation fixes the initial field phase in this periodic problem. The finite conjugate-velocity domain sets the smallest resolved velocity-space scale and must accommodate the transfer of structure by phase mixing over the sampled interval.

The sufficient iteration-convergence condition in Supplementary Eq. (S9) is checked as the field evolves and is most restrictive near the largest amplitude. For the calculations presented here, the time step remains at least one order of magnitude below this limit. Halving the time step and doubling the conjugate-velocity domain changed the saturation amplitude by less than 1%, supporting convergence of this observable. The characteristic update follows the published BOT implementation [23].

Supplementary Fig. S1 tests the numerical solver against the near-threshold Krook result of Breizman et al. [30]. In this regime, the saturation bounce frequency is proportional to the Krook rate, giving a quadratic dependence of the normalized squared bounce frequency on $\beta/(\gamma_L-\gamma_d)$. The calculations reproduce this scaling over $5 \le \beta/(\gamma_L-\gamma_d) \le 8$. Supplementary Methods states the benchmark-specific amplitude normalization explicitly.

The numerical saturation values are approximately 9% above the near-threshold analytical prediction, while reproducing its quadratic trend (Supplementary Fig. S1). This benchmark tests the solver in a known asymptotic regime. The reported refinement check concerns saturation amplitude; the input ordering in Table 2 separately assesses the physical approximation used for the discharge-informed families.

## Particle moments and energy conventions

The change in mean parallel electron energy, denoted $\delta T$, connects distribution reshaping to the energy extracted during the wave event. We define $f(v,t) = \langle F(x,v,t)\rangle_\xi$ as the full phase-averaged energetic-electron distribution and evaluate its second velocity moment [31]:

$$\delta T = \frac{1}{2e} m_e \frac{\int v^2 f dv}{\int f dv} - T_0 \qquad (13)$$

Here $T_0$ is the parallel-energy moment of the initial model distribution, ensuring $\delta T = 0$ initially. In Eq. (13), division by $e$ converts the kinetic-energy moment from joules to eV, and $T_0$ is expressed in the same units. Both moments use the same population normalization and velocity domain. The second moment includes the directed-flow contribution to the parallel energy. The measured discharge temperatures—2.27 eV, 2.52 eV, 2.46 eV, and 2.66 eV at 13.56 MHz, 27 MHz, 37 MHz, and 43 MHz, respectively [18]—parameterize the experimental inputs, while $T_0$ is defined by the initial parallel moment of the model distribution. Thus, $\delta T$ isolates the energy-per-particle response of the modeled energetic population. The current harmonic at the wave number of the single-mode field determines the phase-averaged power transfer. The associated distribution perturbation is

$$\delta F_1(x, v, t) = f_1(v, t)e^{\mathrm{i}\xi} + f_1^*(v, t)e^{-\mathrm{i}\xi}, \tag{14}$$

The zeroth velocity moment gives the density perturbation,

$$\delta n_e = \int_{-\infty}^{+\infty} \delta\, F_1\, \mathrm{d}v, \tag{15}$$

and the first velocity moment gives the particle-flux perturbation,

$$\delta \Gamma_e = \int_{-\infty}^{+\infty} v\ \delta F_1\, \mathrm{d}v, \tag{16}$$

Separating the velocity into its resonant and relative components gives $\delta \Gamma_e = v_r \delta n_e + \int_{-\infty}^{\infty} (v - v_r)\delta F_1 dv$. The first term describes transport of the density perturbation at the resonant velocity; the second measures the velocity moment relative to resonance. Symmetry about the resonant velocity removes the second term, while the density perturbation can retain a finite laboratory-frame flux. This decomposition separates the contributions of phase bunching and velocity-space asymmetry to the coherent particle response. With electron charge $q_e = -e$ and $e > 0$, charge-weighting the particle flux gives the conduction-current perturbation:

$$\delta \mathrm{J}_e = q_e \delta \Gamma_e = -e \delta \Gamma_e, \tag{17}$$

The negative electron charge makes conventional current and electron particle motion oppositely directed. Here $\delta J_e$ is the energetic-electron conduction current. Integrating the kinetic equation over velocity links the density and flux moments directly. With the total distribution vanishing at the velocity boundaries and the spatially uniform Krook rate adopted here, the continuity relation is

$$\frac{\partial \delta n_e}{\partial t} + \frac{\partial \delta \Gamma_e}{\partial x} = -\beta \delta n_e, \tag{18}$$

The right-hand side represents relaxation of the density perturbation through the prescribed Krook reservoir. For complex density and flux coefficients $n_1(t)$ and $\Gamma_1(t)$, the corresponding harmonic relation is $dn_1/dt + (\beta - i\omega)n_1 + ik\Gamma_1 = 0$. This expression retains both envelope evolution and relaxation; the phase-velocity relation $\Gamma_1 \approx (\omega/k)n_1$ follows in the slowly varying, weak-relaxation limit. The local field–particle power density is

$$p_{loc} = \delta E \delta J_e \tag{19}$$

For the full velocity-weighted current, $J_1(t) = -e \int_{-\infty}^{\infty} v f_1(v, t) dv$ and $\delta E = \mathrm{Re}[E_1 \exp(i\xi)]$ give $p_{\mathrm{kin}} = \mathrm{Re}(E_1 J_1^*)$. The one-half prefactor applies when the peak-amplitude current is defined as $2J_1$. The reduced BOT envelope instead uses $J_{res} = -ev_r \int_{-\infty}^{\infty} f_1(v, t) dv$ [20,23] and $p_e = \mathrm{Re}(E_1 J_{\mathrm{res}}^*)$. The difference $p_{\mathrm{kin}} - p_e$ is the contribution from the velocity moment relative to resonance. It vanishes when that moment is neglected consistently in the narrow-resonance closure.

The cold-bulk, single-mode envelope description closes this analysis through the wave-energy balance [13,32]

$$\frac{dW_w}{dt} + p_e + p_d = 0 \tag{20}$$

Here $W_w(t) = (\varepsilon_0 m_e^2 \gamma_L^4/2e^2k^2)|C(t)|^2$ is the phase-averaged mode-energy density, equal to $\varepsilon_0|E_1|^2/2$ in the cold-plasma approximation. The electric-field energy $\varepsilon_0|E_1|^2/4$ and the coherent oscillatory energy of the cold electrons contribute equally. The quantity $p_e$ is the resonant power defined above, and $p_d = 2\gamma_d W_w$ is the background-damping loss. The factor of two in this loss follows because the mode energy is quadratic in an amplitude that damps at rate $\gamma_d$. Integrating Eq. (20) separates temporary wave storage from cumulative damping loss.

Define $K_{res} = \frac{m_e}{2}\int_{-\infty}^{\infty} v^2 f(v,t)dv$ and the prescribed target moment $K_{res,0} = \frac{m_e}{2}\int_{-\infty}^{\infty} v^2 F_0(v)dv$, where $f$ is the phase-averaged distribution used in Eq. (13). The zero-initial-perturbation construction sets $K_{res}(0) = K_{res,0}$. The whole-population kinetic energy obeys $dK_{res}/dt = p_{kin} - \beta(K_{res} - K_{res,0})$, provided the velocity-weighted boundary terms vanish. Equal powers in these two balances correspond to the matched resonant approximation; a full reconstructed current retains their difference. The harmonic current and energy moment must use the same physical distribution scale, integration domain and electron population as the mode normalization. Equation (20) is the reduced wave balance; the Krook reservoir contributes separately to the electron-energy budget. The combined identity below provides a direct consistency test for independently reconstructed moments and wave histories.

## Integrated energy identity

Let $N$ be the constant density of the reconstructed energetic population and $\Delta K = Ne\,\delta T$ its energy-density perturbation. With periodicity and vanishing velocity-boundary fluxes, the phase-averaged number balance is $dN/dt = -\beta(N - N_0)$, where $N_0 = \int_{-\infty}^{\infty} F_0(v)dv$. Thus $N(0) = N_0$ keeps $N$ constant throughout the evolution. Define $W_* = (\varepsilon_0 m_e^2 \gamma_L^4)/(2e^2k^2)$, $Y = |C|^2$, $H = \Delta K/W_*$, $d = \gamma_d/\gamma_L$ and $b = \beta/\gamma_L$. In normalized time, write the electron and mode equations as $H' + bH = Q + R$ and $Y' + 2dY = -Q$, where $Q = p_e/(\gamma_L W_*)$, $R = (p_{\mathrm{kin}} - p_e)/(\gamma_L W_*)$ and a prime denotes differentiation with respect to $\tau$. The correction $R$ retains the in-phase power contribution from the relative-velocity current. A difference between the complex full and resonant currents contributes to $R$ only through its component in phase with the electric field. Eliminating $Q$ and using $H(0) = 0$ gives

$$H(\tau) + Y(\tau) = e^{-b\tau}Y(0) + \int_0^\tau e^{-b(\tau-s)}[R(s) - (2d - b)Y(s)]ds \tag{21}$$

For a matched resonant closure, $R = 0$. With $b = 0$ as well, $-H(\tau) = Y(\tau) - Y(0) + 2d\int_0^\tau Y(s)ds$. Finite recovery introduces an exponential weight on earlier wave activity. When $R = 0$ and $2d \geq b$, non-negativity of $Y$ gives $-H(\tau) \geq Y(\tau) - e^{-b\tau}Y(0)$. The identity and bound apply to a consistent physical reconstruction of the entire modeled population; a finite velocity window requires its boundary terms. Supplementary Methods gives the derivation. The endpoint energy ratios reported in Results are calculated directly from the reported damping-scan values.

## Data availability

The data supporting the reported simulations are available from the corresponding author upon reasonable request. The experimental EEPFs used to construct the model inputs are reported in Ref. [18]. The numerical values used for the ordering map and endpoint energy comparison are given in this article and its Supplementary Information.

## Code availability

The numerical formulation is based on the public BOT code [23]. The production input files, code modifications and diagnostic reconstruction are available from the corresponding author upon reasonable request.

## Acknowledgements

The authors thank Matthew Lilley for the BOT code and for inspiring discussions during Lei Chang's visit at UT Austin. This work was supported by National Natural Science Foundation of China (12675255, 92271113, 12411540222, 12481540165), National Magnetic Confinement Fusion Energy R&D Program of China (2025YFE03070000), Natural Science Foundation Project of Chongqing (CSTB2025NSCQ-GPX0725), Fundamental Research Funds for the Central Universities (2025CDJZKYDYL-02), and ENN's Hydrogen–Boron Fusion Research Fund (2025ENNH B01-011).

## Author contributions

Zhao-Ju Bo: Methodology, Investigation, Data curation, Formal analysis, Validation, Visualization, Writing – original draft; Lei Chang: Funding acquisition, Conceptualization, Writing – review & editing, Resources, Supervision; Shi-Jie Zhang: Visualization, Writing – review & editing; Ying-Xin Zhao: Writing – review & editing, Visualization; Zi-Chen Kan: Visualization, Writing – review & editing; Ji-Kai Sun: Writing – review & editing; Ilya Zadiriev: Writing – review & editing; Elena Kralkina: Writing – review & editing; Shin-Jae You: Writing – review & editing; Jing-Jing Ma: Writing – review & editing; Husi Géza: Writing – review & editing.

## Competing interests

The authors declare no competing interests.